\documentclass[prl,superscriptaddress,twocolumn,showpacs,preprintnumbers,amsmath,amssymb]{revtex4}

\usepackage[dvips]{graphicx}
\usepackage{dcolumn}
\usepackage{bm}

\begin{document}


\title{Electronic Structure of Charge- and Spin-controlled Sr$_{1-(x+y)}$La$_{x+y}$Ti$_{1-x}$Cr$_x$O$_3$}

\author{H. Iwasawa}
\affiliation{Department of Applied Physics, Tokyo University of Science, Shinjuku-ku, Tokyo 162-8601, Japan}
\affiliation{National Institute of Advanced Industrial Science and Technology, Tsukuba, Ibaraki 305-8568, Japan}

\author{K. Yamakawa}
\author{T. Saitoh}
	\email[Author to whom correspondence should be addressed.\\ Electronic address: ]{t-saitoh@rs.kagu.tus.ac.jp}
\affiliation{Department of Applied Physics, Tokyo University of Science, Shinjuku-ku, Tokyo 162-8601, Japan}

\author{J. Inaba}
\affiliation{Department of Physics, Waseda University, Tokyo 169-8555, Japan}

\author{T. Katsufuji}
\affiliation{Department of Physics, Waseda University, Tokyo 169-8555, Japan}
\affiliation{PRESTO, Japan Science and Technology Corporation, Saitama 332-0012, Japan}

\author{M. Higashiguchi}
\affiliation{Graduate School of Science, Hiroshima University, Higashi-Hiroshima 739-8526, Japan}

\author{K. Shimada}
\affiliation{Hiroshima Synchrotron Radiation Center, Hiroshima University, Higashi-Hiroshima 739-8526, Japan}

\author{H. Namatame}
\affiliation{Hiroshima Synchrotron Radiation Center, Hiroshima University, Higashi-Hiroshima 739-8526, Japan}

\author{M. Taniguchi}
\affiliation{Graduate School of Science, Hiroshima University, Higashi-Hiroshima 739-8526, Japan}

\date{\today}

\begin{abstract}

We present the electronic structure of Sr$_{1-(x+y)}$La$_{x+y}$Ti$_{1-x}$Cr$_x$O$_3$ investigated by high-resolution photoemission spectroscopy.
In the vicinity of Fermi level, it was found that the electronic structure were composed of a Cr 3$d$ local state with the $t_{2g}^{3}$ configuration and a Ti 3$d$ itinerant state.
The energy levels of these Cr and Ti 3$d$ states are well interpreted by the difference of the charge-transfer energy of both ions.
The spectral weight of the Cr 3$d$ state is completely proportional to the spin concentration $x$ irrespective of the carrier concentration $y$, indicating that the spin density can be controlled by $x$ as desired. 
In contrast, the spectral weight of the Ti 3$d$ state is not proportional to $y$, depending on the amount of Cr doping.

\end{abstract}


\maketitle


If the charge and spin degrees of freedom of electrons can be controlled just as one designed, it is possible to engineer a next-generation of devices merging conventional electronics with magnetoelectronics \cite{Wolf01}.
So far, however, there is no report that has succeeded in handling magnetic and semiconducting properties independently.
Magnetic semiconductors (MSs) are one of the best candidates among the spin-electronic devices utilizing the both properties.
Indeed, following a successful synthesis of the prototypical In$_{1-x}$Mn$_x$As and Ga$_{1-x}$Mn$_x$As \cite{Munekata89,H.Ohno98,Y.Ohno99}, several room temperature (RT) ferromagnetic MSs such as ZrTe:Cr \cite{Saito03} have recently been discovered, leading a continuous effort to understand the electronic and magnetic properties of these materials.
However, it is still difficult to control their magnetic and semiconducting properties independently, and this difficulty stems from that the both properties depend upon only the one-transition-metal (1-TM) element doping in either III-V-based or RT ferromagnetic MSs.

Alternatively, a Cr-doped perovskite-type titanate Sr$_{1-(x+y)}$La$_{x+y}$Ti$_{1-x}$Cr$_x$O$_3$ (SLTCO) is based on a new concept of independent control of the charge and spin degrees of freedom by 2-TM elements (Ti and Cr); the end compound SrTiO$_3$ of this family is a band insulator with a wide band gap of 3.2 eV.
It is well known that carrier doping can be realized by La$^{3+}$ substitution for Sr$^{2+}$, which introduces electrons ($y$) into Ti 3$d$ conduction band.
A further manipulation of the resulting Sr$_{1-y}$La$_y$TiO$_3$ is to introduce Cr$^{3+}$ by replacing SrTiO$_3$ with LaCrO$_3$ by the amount of $x$.
Because Cr$^{3+}$ (3$d^3$) ions are usually magnetic in oxides, the above substitution of Cr$^{3+}$ for Ti$^{4+}$ would realize a ``spin doping" of S=3/2 local moment at the Ti sites.
In other words, $x$ and $y$ nominally represent ``spin" and ``carrier" concentration ($n_s$ and $n_c$), respectively.
In fact, magnetic and transport measurements showed that the paramagnetic Curie constant increases with $x$, and a semiconducting resistivity decreases with increasing $y$ \cite{Inaba05}.
Judging from the sign change of the Weiss temperature, the correlation between Cr ions should vary from antiferromagnetic to ferromagnetic interaction with from $y$=0 to $y$$>$0.
This ferromagetic interaction is probably caused by the exchange interaction between the Cr 3$d$ (spin) and Ti 3$d$ (carrier) electrons, but the paramagnetic Curie temperature ($T_{\rm c}$) of this system is still low below $\sim$10 K unfortunately.
Nevertheless, it is not evident that the doped Cr$^{3+}$ really form the local spin of S=3/2 since both Ti and Cr $t_{2g}$ orbitals will be partially filled.
In this sense, despite the low $T_{\rm c}$, SLTCO is worthy to be investigated as a prototype of next-generation MSs in which the two degrees of freedom can be controlled.

The primary focus of this Letter is to elucidate the electronic structure of SLTCO as well as a possibility of an independent control of charge and spin degrees of freedom by high-resolution photoemission spectroscopy (PES).
The result demonstrates that the near-$E_{\rm F}$ electronic structure consist of the Cr 3$d^3$ local state and Ti 3$d$ itinerant state.
We will show that the relative energy levels of these 3$d$ states can be described by the difference of the effective charge-transfer energy ($\Delta_{\rm eff}$) of both ions, which is one of the basic parameters in the electronic structure of conventional 1-TM perovskite-type oxides \cite{Zaanen85}.
We discuss that an appropriate choice of two TM elements can realize a new magnetic system in which local spins and charge carriers are independently controlled.


Polycrystalline samples of Sr$_{1-(x+y)}$La$_{x+y}$Ti$_{1-x}$Cr$_x$O$_3$ with ($x$, $y$)=(0.1, 0.1), (0.2, 0.1), (0.2, 0.2), and (0.2, 0.15) were prepared by solid-state reaction \cite{Inaba05}.
We will simply denote them as (0.1, 0.1), etc. 
All the measurements were carried out at BL-1 of Hiroshima Synchrotron Radiation Center (HSRC), Hiroshima University using a SCIENTA ESCA200 electron analyzer \cite{Shimada01, Shimada02}. 
In order to obtain clean surface, we fractured the samples {\em in situ} in ultrahigh vacuum (better than 1$\times$10$^{-10}$ Torr) at 50 K.
The spectra were taken at 50 K using selected photon-energies (45.6, 50.0, and 150 eV), and the overall experimental energy resolution was set at 50 meV.
The backgrounds due to the unscattered electrons were subtracted from all the spectra.
For comparison, a spectrum of Sr$_{1-(x+y)}$La$_{x+y}$Ti$_{1-x}$V$_x$O$_3$ (SLTVO) with (0.2, 0.2) was also recorded under the same experimental conditions as above.


\begin{figure}[t]
	\begin{center}
	\includegraphics[width=70mm,keepaspectratio]{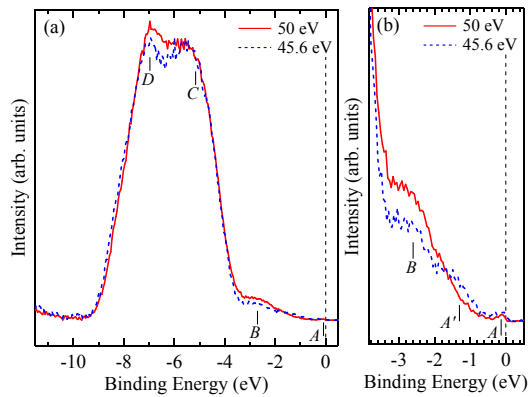}
	\end{center}
	\caption{(Color online). Resonant PES spectra of (0.2, 0.15) in (a) the full valence-band and (b) the near-$E_{\rm F}$ region. The spectra are taken at 45.6 eV (dashed line) and 50.0 eV (solid line), which correspond to the Ti and Cr 3$p$$-$3$d$ on resonance, respectively.}
\label{FIG1}
\end{figure}

In order to identify the Ti and Cr 3$d$ states, we have first performed a resonant PES measurement for (0.2, 0.15).
Figure~\ref{FIG1} shows the resonant PES spectra of (0.2, 0.15) taken at 45.6 eV (dashed line) and 50.0 eV (solid line), corresponding to the Ti and Cr 3$p-$3$d$ on resonance, respectively.
In Fig.~\ref{FIG1} (a), one can observe the four spectral features denoted as $A$ to $D$.
By comparison with the reported PES spectra of La$_x$Sr$_{1-x}$TiO$_3$ \cite{Fujimori92,Aiura02,Yoshida02}, the leading feature $A$ (-0.7 eV) is attributed to the Ti 3$d$ state, and the dominant features $C$ (-5.1 eV) and $D$ (-6.9 eV) have primarily the O 2$p$ character. 
Consequently, the remaining feature $B$ (-2.7 eV) is the most likely Cr 3$d$ state.
These assignments are confirmed by the resonant enhancement of the features $A$ and $B$ at each absorption threshold as shown in Fig.~\ref{FIG1} (b).
Note that the Ti 3$d$ state further split into the quasiparticle band and the remnant of the lower Hubbard band corresponding to the one-electron excitation from $d^1$ to $d^0$ of the insulating LaTiO$_3$, namely, the coherent ($A$) and incoherent ($A'$) parts, consistent with the literature \cite{Yoshida02}.  

\begin{figure*}
	\begin{center}
	\includegraphics[width=140mm,keepaspectratio]{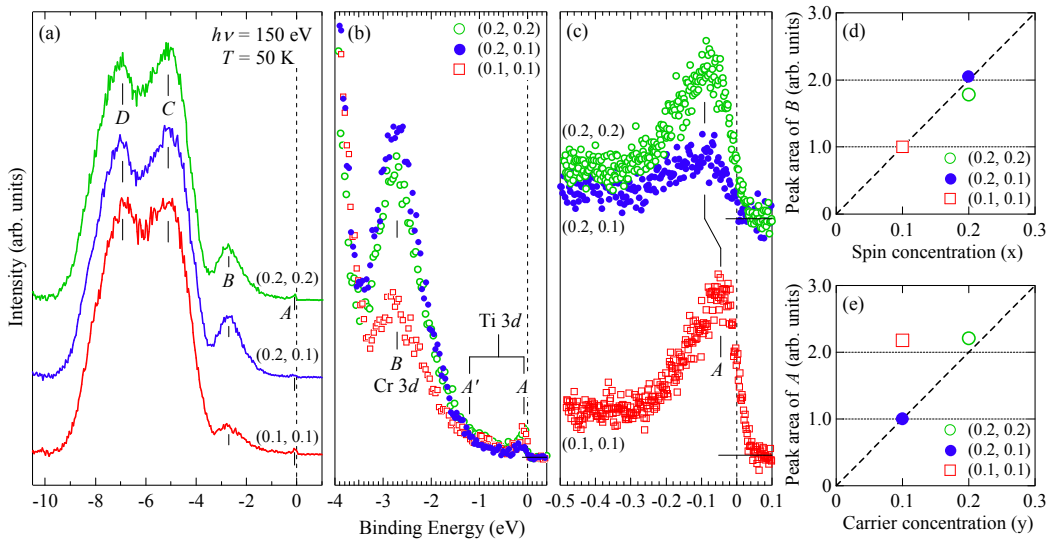}
	\end{center}
	\caption{(Color). (a) Valence-band spectra taken at 150 eV. (b) Same as (a) in the near-$E_{\rm F}$ region in an expanded scale. (c) Very near-$E_{\rm F}$ region spectra. (d) and (e) Intensity plots of the features $B$ and $A$, respectively. Spectral weight of features $B$ and $A$ is set to unity at (0.1, 0.1) and (0.2, 0.1). The integration windows are given in the text.} 
\label{FIG2}
\end{figure*}

To see how the charge and spin degrees of freedom are controlled, we have investigated a doping dependence of PES spectra.
Panel (a) of Fig.~\ref{FIG2} presents the valence-band spectra of SLTCO for several compositions taken at 150 eV. 
At first sight, the four spectral features $A-D$ identified in Fig.~\ref{FIG1} (a) can be observed.
Owing to the larger photoionization cross section of the Ti and Cr 3$d$ states relative to one of the O 2$p$ state, the near-$E_{\rm F}$ structures are much clearer and sharper than in Fig.~\ref{FIG1} (a).
An apparent doping dependence of the features $B$ and $A$ can be seen easily in the near-$E_{\rm F}$ [panel (b)] and the very near-$E_{\rm F}$ [panel (c)] region spectra, respectively.
Panels (d) and (e) summarize this doping dependence, where the integration windows are from -3.3 to -2.1 eV ($B$) and from -0.4 to 0.1 eV ($A$). 

We can find two important facts about Cr doping from the figures.
First, panels (a) and (b) clearly show that the feature $B$ is located at a deep level far below $E_{\rm F}$ (${\sim}$2.7 eV).
This is demonstrating that the doped Cr 3$d$ electrons do not participate in electric conduction.
Second, it is likely that the doped Cr ions are always trivalent with the $t_{2g}^3$ local configuration because the peak shift of the Cr 3$d$ and O 2$p$ bands due to the doping of both $x$ and $y$ is not observed in the spectra.
This is contrary to the previous PES spectra of La$_{1-x}$Sr$_x$CrO$_3$, in which the Cr 3$d$ peak and O 2$p$ bands show systematic changes with increasing Cr$^{4+}$ localized state by replacing La$^{3+}$ for Sr$^{2+}$ \cite{Maiti96}. 
Here it may be still possible that they become all tetravalent, although such a situation is very unlikely.
However, this can be ruled out from the result of panels (d) and (e); panel (d) shows that the spectral weight of Cr 3$d$ state is completely proportional to $x$ irrespective of $y$, indicating that the Cr ions do not change their valence.
Panel (e) also shows that spectral weight of coherent part has a linear relationship with $y$ [except for (0.1, 0.1), which will be discussed later].
If the Cr ions are tetravalent, ($x$, $y$)=(0.2, 0.1) and (0.2, 0.2) mean $n_c$=0.2+0.1 and 0.2+0.2, respectively.
This is, however, resulting in the spectral weight ratio of 0.4/0.3$\approx$1.33 between the two samples, which is far from the observed linear relationship. 
Therefore, we believe that the doped Cr ions form the S=3/2 ($t_{2g}^3$) local-spin state and $x$ gives $n_s$ of this local spin, although a more precise determination would be done using core-level PES.

On the other hand, panel (e) shows that the spectral weight of (0.1, 0.1) deviates from the linear relationship.
This indicates that the charge carrier is not controlled solely by $y$.
In other words, the density of states near $E_{\rm F}$ decreases with increasing Cr doping despite the same carrier concentration.
This is in perfectly agreement with the electrical resistivity data \cite{Inaba05}.
Moreover, the peak position of the feature $A$ apparently shifts away from $E_{\rm F}$ by $\sim$50 meV upon Cr doping without changing $y$ as shown in panel (c).
This is, hence, not a rigid-band shift and what is observed is actually the shift of the midpoint of the leading edge and a strong spectral weight suppression, both of which are signals of opening a pseudogap.
However, since the lineshape of the feature $A$ is still quasiparticle-like, the charge carriers are not completely localized and the system is very in the vicinity of the metal-insulator transition.
Under such circumstances, the long-range Coulomb interaction becomes dominant and if a sufficient potential disorder exists, the well-known Coulomb gap will appear \cite{Efros75,Davies86,Sarma98}.
In our case, the Cr doping $x$ should give rise to a strong potential disorder at Ti sites with a small number of the charge carriers $y$.
Hence we believe that this pseudogap is most likely the Coulomb gap.  
Further investigation is needed on this issue, but it is beyond the scope of this Letter.


\begin{figure}
	\begin{center}
	\includegraphics[width=77mm,keepaspectratio]{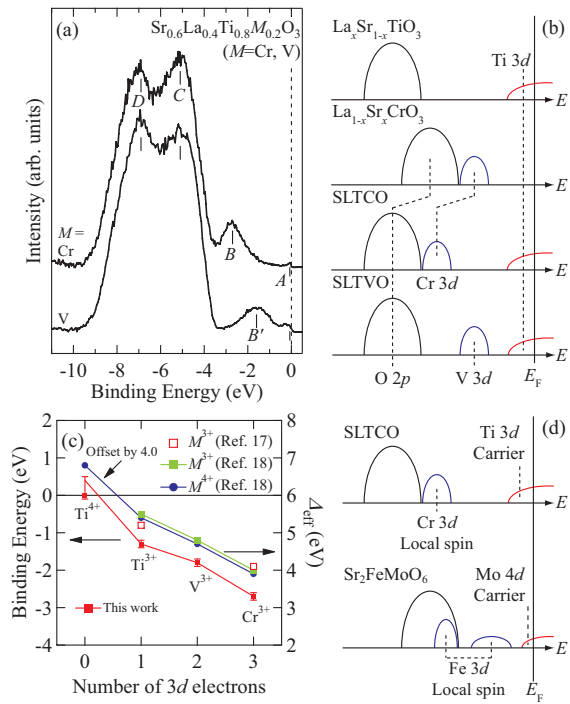}
	\end{center}
	\caption{(Color). (a) Valence-band spectra of Sr$_{0.6}$La$_{0.4}$Ti$_{0.8}$$M$$_{0.2}$O$_3$ ($M$=Cr, V) taken at 150 eV. (b) and (d) Schematic electronic structure of SLTCO compared with the analogues and the double perovskite Sr$_2$FeMoO$_6$. (c) Binding energy of spectral features and $\Delta$$_{\rm eff}$'s are plotted against the number of 3$d$ electrons, where $\Delta$$_{\rm eff}$ for $M^{4+}$ includes an offset by 4.0 eV. Since the centroid of $A$ must be above $E_{\rm F}$, the upper error bar of the Ti$^{4+}$ is expected to be large.}
\label{FIG3}
\end{figure}

Finally, we discuss how $\Delta_{\rm eff}$ of TM elements plays a crucial role in the electronic structure of the 2-TM systems.
Panel (a) of Fig.~\ref{FIG3} shows the valence-band spectra of the Cr- and V-doped titanates Sr$_{0.6}$La$_{0.4}$Ti$_{0.8}$$M_{0.2}$O$_3$ ($M$=Cr, V).
Comparing the two spectra, one can easily attribute the feature $B'$ (-1.8 eV) to the V 3$d$ state.
Panel (b) displays a schematic electronic structure of the four compounds deduced from our results and the literature \cite{Aiura02, Yoshida02, Maiti96}.
The distance between the Cr 3$d$ state and the O 2$p$ band in SLTCO is virtually identical to that in La$_{1-x}$Sr$_x$CrO$_3$ \cite{Maiti96}.
On the other hand, the Cr 3$d$ peak in SLTCO is located at a deeper energy level than in La$_{1-x}$Sr$_x$CrO$_3$ because the $E_{\rm F}$ location is determined by Ti 3$d$ state.
In other words, the combination of (1) the difference of the $\Delta_{\rm eff}$'s ($D$$\Delta_{\rm eff}$) of the Ti and Cr 3$d$ states and (2) the $E_{\rm F}$ location determined by the Ti 3$d$ state realizes the Cr 3$d$ localized state (and hence the local spins) in the band gap between O 2$p$ and Ti 3$d$ bands in SLTCO.

This is quantitatively demonstrated in panel (c), in which the binding energy of the observed spectral features is compared with $\Delta_{\rm eff}$'s deduced from the analysis of PES spectra for $M^{3+}$ \cite{Tom95} and from the empirical relation for $M^{3+}$ and $M^{4+}$ \cite{Fujimori93}.
The panel reveals that an appropriate offset for $\Delta_{\rm eff}$ for $M^{3+}$ or $M^{4+}$ reproduces the relative energy of these features very well \cite{comment1}.
As a consequence, the panel predicts that V ions in SLTVO is not V$^{4+}$ but V$^{3+}$ because the $B$' just locates between Ti$^{3+}$ and Cr$^{3+}$.
It should be noted that the electronic structure is not solely determined by $D\Delta_{\rm eff}$ or $\Delta_{\rm eff}$ but also by the effective O 2$p$-TM $d$ hybridization ($T_{\rm eff}$) and the effective on-site $d$-$d$ Coulomb interaction $U_{\rm eff}$ \cite{comment2}.
Indeed, ($\Delta_{\rm eff}$-$U_{\rm eff}$) is observed directly in the PES spectra instead of $\Delta_{\rm eff}$.
However, the effects of $U_{\rm eff}$ and $T_{\rm eff}$ do not appear in the energy {\em difference} among the features because $U_{\rm eff}$ and $T_{\rm eff}$ change systematically with number of $d$ electrons \cite{comment3}.
This is why the relative binding energy of the features directly reflects $D\Delta_{\rm eff}$.
Therefore, $D\Delta_{\rm eff}$ is an essential parameter to determine the backbone of the electronic structure of 2-TM compounds.

Owing to the large ($\sim$2.7 eV) $D\Delta_{\rm eff}$, SLTCO has local spins and charge carriers.
This indicates that one can introduce a localized spin at a deep energy level by an appropriate choice of two elements with a relatively large $D\Delta_{\rm eff}$. 
In addition, such a situation should be realized not only in the above randomly-substituted 2-TM compounds, but also in the ordered 2-TM compounds, namely, the ordered double perovskite-type oxides.
Panel (d) compares the schematic electronic structure of SLTCO and a typical half-metallic double perovskite Sr$_2$FeMoO$_6$ \cite{Tom02}.
In the case of Sr$_2$FeMoO$_6$, the location of $E_{\rm F}$ is determined by the Mo 4$d$ (+Fe 3$d$) $t_{2g\downarrow}$ state and the Fe 3$d^5$ $t_{2g\uparrow}^{3}e_{g\uparrow}^{2}$ state are located between the O 2$p$ and Mo 4$d$ states, reflecting a large $D\Delta_{\rm eff}$ between Fe$^{3+}$ and Mo$^{5+}$.
That is to say, the Ti and Cr 3$d$ states correspond to the Fe 3$d$ and Mo 4$d$ states, although there exists the essential difference between them, $n_s$-to-$n_c$ ($x$-to-$y$) ratio and the randomness.
This comparison further implies that SLTCO could be double-exchange type ferromagnet like La$_{1-x}$Sr$_x$MnO$_3$ \cite{review}.

Although it is simplified, our interpretation employing $D\Delta_{\rm eff}$ gives a useful view for understanding the electronic structure of 2-TM systems as well as new-material designing. 
For instance, it is interesting to apply it to the electronic structure of interface layers, which has hardly been studied by PES in spite of its importance for industrial applications \cite{Kumi04}. 
In the heterointerfaces, the electronic structure should be considered as quasi-two-TM system, and hence it is essential to take account of the $\Delta_{\rm eff}$'s of the two TM ions in each thin film.
This would give essential information about the electronic structure of heterointerfaces, particularly on the amount of charge transfer from one to the other \cite{Kumi04}.


In conclusion, we have studied the electronic structure of Sr$_{1-(x+y)}$La$_{x+y}$Ti$_{1-x}$Cr$_x$O$_3$ by high-resolution photoemission spectroscopy.
Near-$E_{\rm F}$ electronic structures show two-component structure of the Cr 3$d$ local state and Ti 3$d$ itinerant state.
Owing to the large $D\Delta_{\rm eff}$ of Cr and Ti ions, the doped Cr 3$d$ electrons realize S=3/2 ($t_{2g}^3$) local-spin state at a deep level far below $E_{\rm F}$ ($\sim$2.7 eV). 
An appropriate choice of 2-TM elements provides an opportunity to create a new magnetic system in which local spins and charge carriers are independently controlled.

We would like to thank S. Kaneyoshi for useful discussions.
This work was partly supported by a Grant-in-Aid for COE Research (13CE2002) by MEXT of Japan.
The synchrotron radiation experiments have been done under the approval of HSRC (Proposal Nos. 03-A-58 and 04-A-39).


\begin{thebibliography}{99}
\bibitem{comment1}Because the electron doping yields the $N$- and ($N$+1)-electron degenerated ground state, we can compare $D\Delta_{\rm eff}$ of the $N$-electron system with the ($N$-1) photoemission final states.
\bibitem{comment2}The Hund's rule energy stabilization also plays an important role in the compounds which have very large local spins.
\bibitem{comment3}Not only $\Delta_{\rm eff}$ but also $U_{\rm eff}$ must be taken into account for the $d^5$ configuration because both become largest due to the large positive multiplet correction.
\end{thebibliography}

\end{document}